\documentclass[aps,showpacs,nofootinbib]{revtex4}
\usepackage{epsf,latexsym}

\begin{document}

\title{A proof of the Bekenstein bound for any strength of gravity 
through holography}
\author{Alessandro Pesci}
\email{pesci@bo.infn.it}
\affiliation
{INFN-Bologna, Via Irnerio 46, I-40126 Bologna, Italy}

\begin{abstract}
The universal entropy bound of Bekenstein is considered,
at any strength of the gravitational interaction.
A proof of it is given, 
provided the considered general-relativistic spacetimes allow for
a meaningful and inequivocal definition of the quantities
which partecipate to the bound 
(such as system's energy and radius).
This is done
assuming as starting point that,
for assigned statistical-mechanical local conditions,
a lower-limiting scale $l^*$
to system's size definitely exists,
being it required by holography
through its semiclassical formulation
as given by the
generalized covariant entropy bound.

An attempt is made also to draw
some possible general consequences of the $l^*$ assumption
with regards to the proliferation of species problem and
to the viscosity to entropy density ratio. 
Concerning the latter, various fluids are considered
including systems potentially relevant, to some extent,
to the quark-gluon plasma case.
\end{abstract}


\maketitle

$ $
\section{Motivation}
In the description of thermodynamic systems,
recent works \cite{Pesci2,Pesci3} 
have shown that the generalized covariant entropy bound (GCEB) \cite{FMW},
which can be considered as the most general formulation
of the holographic principle for semiclassical circumstances,
is universally satisfied if and only if
the statistical-mechanical description
is characterized by
a lower-limiting spatial scale $l^*$, 
determined by the assigned 
local thermodynamic conditions.\footnote{Given that the GCEB implies
the generalized second law (assuming the validity
of the ordinary second law) \cite{FMW},
a support to, or an echo of, the statement above 
can be found in \cite{FBB},
where,
for an ideal fluid accreting onto a black hole,
if no lower-limit is put
to the spatial scale of the thermodynamic description
of the fluid,
a violation of the generalized second law is obtained.}
In \cite{Pesci4} the consequence has been drawn
that $l^*$ entails also a lower-limit to the temporal scale,
and this supports, 
both for its existence and value,
the recently proposed universal bound \cite{Hod} 
to relaxation times of perturbed thermodynamic systems,
bound attained by black holes.\footnote{An 
expression for the actual limit 
to relaxation times of statistical systems
(the existence of a limit
was broadly expected 
from general quantum arguments \cite{Landau})
was first given in \cite{Sachdev1}
through explicit consideration of relaxation times of 
systems at quantum-critical conditions.
A recent account of the intriguing underlying connection
between critical (flat-spacetime) quantum statistical physics
and black hole thermodynamics can be found in \cite{SachdevMuller}.} 
All this amounts to say that
the GCEB is satisfied 
if and only if a fundamental discreteness is present
in the spatio-temporal description of statistical-mechanical systems,
with a value changing from point to point
being determined by local thermodynamic conditions. 

The exact value of $l^*$
is set from the GCEB \cite{Pesci2,Pesci3}.
This does not necessarily mean however 
that $l^*$ has to depend on gravity. In effect, the derivation
of $l^*$ from the GCEB is through
Raychaudhuri equation so that gravity cancels out \cite{Pesci3}.
Indeed, this scale appears intrinsically unrelated to gravity
(in particular $l^*$ must not be thought as something around the Planck scale;
it is in general much larger than this) 
since, as we will discuss later,  
it can be fully described simply as a consequence
or expression of the flat-space quantum description of matter \cite{Pesci3}.

Among the various entropy bounds, the GCEB 
appears the most general one,
subsuming in some way all other previous bounds
at the conditions in which they are supposed to hold.
The Bekenstein universal entropy
bound (UEB) \cite{Bekenstein1}, 
the first proposed among them, 
has been shown
to be implied by GCEB for weak field conditions
(a strengthened version of it, actually, in case 
of non-spherical systems) \cite{Bousso2003}.
This bound sets an upper limit to the $S/E$ ratio
($S$ is entropy and $E$ is total mass-energy)
for a given arbitrary system, in terms of its size.
The original motivation was a {\it gedanken} experiment,
in the context of black hole physics,
in which
the variations of total entropy are considered when
objects with negligible self-gravity
are deposited at the horizon.
In spite of how it was derived,
it has been clear since the beginning 
that gravity should not play any role 
in establishing the UEB \cite{Bekenstein1}.
Since,
besides the systems with negligible self-gravity,
also the objects with the strongest gravitational effects,
namely the black holes, appear to satisfy the bound
(actually they attain it),
the original proposal stressed moreover
the UEB should hold true for whatever strength
of the gravitational interaction (assuming the conditions
be such that the terms entering
the bound can be appropriately defined).
A strong support, if not a proof, of this statement
under safe conditions for the definition of system's ``energy'' and ``size'',
has been given in \cite{SWZ}, through the consideration
of self-gravitating radiation confined in a spherical box;
in particular, some key elements have been pointed out there,
apparently crucial in determining the validity of
the UEB under strong self-gravity. 

What is left is the intriguing situation
of a bound belonging in principle
to pure flat-spacetime physics,
which however should hold true, exactly with the same form,
also under conditions in which 
gravity is strong.
The main goal of present work
is just trying to face again this issue,
this time from a holographic standpoint.
This is done
giving a general argument supposed to prove
the UEB for any strength of gravity
(at suitable conditions, 
i.e. with all the terms involved in the bound
unequivocally defined) as a result of 
the existence and value of the limiting scale $l^*$,
that is from holography
in its semiclassical formulation.
We then explore what
this holographic perspective says about
the connection of the UEB with the $\eta/s$ 
Kovtun-Son-Starinets bound (KSS) \cite{KSS}
and with the proliferation of species problem.

\section{The limiting length}\label{limiting}

Given a spacelike 2-surface $B$
the covariant entropy bound \cite{Bousso,Bousso_review}
is a conjecture that
connects the entropy content of adjacent regions
to the area $A$ of $B$.
It states that $S(L) \leq A/4$
(in Planck units, the units we will use throughout this paper)
where $L$ is a lightsheet of $A$
(a 3d null hypersurface generated
by not expanding light rays orthogonal to $B$
and followed until a caustic or a singularity is reached)
and $S(L)$ is the entropy on it.
The generalization in \cite{FMW}, i.e.  the GCEB,
is in that the lightsheet $L$ is allowed to terminate
on a second spacelike 2-surface with area $A^\prime$,
before the caustic or singularity are reached,
with the bound becoming $S(L) \leq (A-A^\prime)/4$.

The proof of the GCEB for perfect fluids given in \cite{Pesci2}
(other proofs, in terms of conditions 
which are proven
to suffice
for the validity of the bound,
are in \cite{FMW, Proof})
has allowed to identify a condition
not only sufficient 
but also necessary 
for the validity of the GCEB 
\cite{Pesci2,Pesci3}. 
This condition,
which turns out to be in relation
with the condition (1.9) in \cite{FMW},
has been expressed \cite{Pesci3} in terms of
a lower-limiting spatial length $l^*$ to the size
of any system, 
for which an assigned statistical-mechanical description
is given.
This amounts to put everywhere 
a local limit to
the size $l$ of the disjoint sub-systems
of which a large system can be thought as composed of.
The condition is \cite{Pesci3} 

\begin{eqnarray}\label{condition}
l \geq l^*,
\end{eqnarray}
with, for a perfect fluid, 

\begin{eqnarray}\label{lstar}
l^* = \frac{1}{\pi} \ \frac{s}{\rho + p}
= \frac{1}{\pi T} \ \left( 1 - \frac{\mu n}{\rho + p} \right),
\end{eqnarray}
being for thin slices the GCEB saturated
when their thickness is as small as $l^*$ (when feasible).
Here 
$s$, $\rho$, $n$, $p$, $T$, $\mu$ are respectively
local entropy, mass-energy and number densities,
local pressure, temperature and chemical potential (where the latter 
includes the rest energy, if any, of constituent particles).

The need for such limiting scale $l^*$
can be recognized through consideration of thin plane slices.
The variation of the cross-sectional area of the lightsheet
we trivially can construct on them is quadratic in the thickness $l$ 
of the layer,
whereas its entropy content is (obviously) linear in $l$.
Hence, 
if a lower limit is not envisaged for $l$,
for thin enough (terminated) lightsheets the GCEB would be
definitely violated.

Looking at a system consisting of a single thin slice
of thickness $l$ and area $A$,
condition (\ref{condition}) amounts to say
that, for $l < l^*$, 
a statistical-mechanical description according to which
the system has a pressure $p$
and a proper energy $E$ and entropy $S$ 
(to give $\rho = E/A l$ and $s = S/A l$)
matching the assigned values which define $l^*$, 
cannot be given,
that is, the size of the layer somehow forbids that the thermodynamic
potentials have the values assigned.
One important consequence of this
is that
the scale associated
to local thermal equilibrium at any point
cannot be lower than $l^*$ there.
The concept of local thermal equilibrium scale $l_{eq}$
appears, in fact, such that if slices of thickness $l_{eq}$
are physically cutted, the intensive thermodynamic potentials
in them have to remain unchanged
(that is they would have negligible boundary effects),
and this if $l_{eq} < l^*$ would imply a violation of the above.
Indeed, in general the scale of thermal equilibrium 
can be expected to be
much larger than $l^*$. This is implied
by the notion itself of $l^*$:
for macroscopic systems at global equilibrium, for example,
the scale of thermal equilibrium is given by system's size,
which can be large at will,
whereas $l^*$ remains fixed, 
as determined by the assigned local thermodynamic parameters.

Also,
we have no reason to expect that the 
unallowance of an assigned statistical-mechanical
description should set in exactly at the scale $l^*$;
we should expect instead that in general there will be a gap
between $l^*$ and the lowest scale $l_{alw}$ allowed by the material medium,
being the gap dependent on the properties of the latter 
(after all, we know
that there are fluids much more entropic than others
so that, 
for the less entropic fluids, 
both the GCEB for any configuration 
and thus also the inequality (\ref{condition}),
have to be satisfied to spare).
This implies that a mechanism must be at work,
which, even putting $l^*$ and the GCEB aside completely, fixes $l_{alw}$.

This mechanism is quite naturally provided by the uncertainty principle.
If we ideally
slice up a macroscopic system with finer and finer spacing $l$,
we can expect to obtain thinner and thinner slices inside a region of
thermal equilibrium, each of them with always a same 
statistical-mechanical description.
For point-like constituents
(i.e. particles or molecules for which the intrinsic 
spatial quantum uncertainty is larger than their size
as composite objects), 
this will remain true
down to a certain limiting thickness
of the slices $l_{qm}$,
driven by quantum mechanics and 
roughly corresponding to the representative 
de Broglie wavelength $\lambda$
of the constituent particles,
at which scale
the statistical-mechanical properties of
the material medium (such as temperature and energy density for example)
begin to be affected by slicing, if physical.
It is not possible, for example, consider systems consisting
of a slice of a photon gas at temperature $T$ with thickness
$l < l_{qm} = \frac{1}{\pi T}$ \cite{Pesci3}.
Put another way,
a statistical-mechanical description (with boundary effects included)
for 3d containers, appears quantum-mechanically untenable
when one of the dimensions becomes $l \ll \lambda$, if
$\lambda$ is the pretended particle's wavelength.

For a system consisting of particles with a same
wavelength $\lambda$, this suggests $l_{qm} = \lambda$.
In the general case we still will write $l_{qm} = \lambda$
with $\lambda$ representing a typical or characteristic wavelength
of constituent particles, meaning that below $\lambda$
the quantum uncertainty from the boundary starts to affect
the intrinsic thermal distribution of the particles in
the system.

For `virtual' or `mathematical' slicing,
for $l < l_{qm}$ particles `in' a slice affect also adjacent slices,
so that extensivity is lost in the sense that entropy
or energy cannot be recovered as sum on slices of thickness 
$l < l_{qm}$. 
The entropies or energies of these slices
must instead be combined
in a more complicated manner,
to give however always those same values on scales $l > l_{qm}$.
The net effect is thus anyhow that total entropy or energy are determined
as sum on slices, each not thinner than $l_{qm} = \lambda$. 
 
In this sense we thus can never go below $\lambda$ 
and, obviously, below the size $l_c$
of the constituent particles due to compositeness,
if dominant.
Moreover, for point-like constituents the limiting scale $\lambda$
can always be attained.
In case of composite constituents 
we should stop at $l_c \geq \lambda$,
but we can always imagine an equivalent
system composed of particles with mass and other properties
equal to those
in the original system but point-like, so that
the limiting scale $\lambda$ can still be reached.

Thus, 
$\lambda$ (or the maximum between $\lambda$ and $l_c$)
could possibly play the role of $l_{alw}$ above.
However,
how confronts $\lambda$ with $l^*$ 
as given by (\ref{lstar}) ?
After all, if it could be $\lambda < l^*$ for some system
with point-like constituents, the GCEB could be violated
(choosing slices
with thickness as small as $\lambda$).
The consideration of various statistical-mechanical systems
with point-like constituents hints that 

\begin{eqnarray}\label{condition2}
\lambda \geq l^*,
\end{eqnarray} 
with in general $\lambda \gg l^*$,
being the bound attained
for ultra-relativistic systems (with $\mu = 0$) \cite{Pesci2,Pesci3}.
Uncertainty principle seems to conspire in order that
the limit in (\ref{condition2})
be strictly attained in the most challenging cases
and, as a consequence, 
be satisfied to spare by more ordinary systems.
According to this,
the conjecture can be made that inequality (\ref{condition2})
is guaranteed and exactly required by quantum mechanics.
In this perspective
inequality (\ref{condition}), 
namely the GCEB 
when the inequality is taken in a general-relativistic context, 
precisely predicts and is predicted 
by quantum mechanics.\footnote{Through the consideration of 
single-particle systems,
in \cite{Bousso2003, Bousso2004} the role,
under weak field conditions,
of the UEB
(or, at these conditions, of the GCEB)
in predicting the uncertainty principle
has been stressed.}
In any case,
even with this conjecture aside,
GCEB + quantum mechanics
means $l^* \leq \lambda \leq l_{alw}$ universally.

All the above says that
an upper-bound to the
quantity $\frac{s}{\rho + p}$
given by

\begin{eqnarray}\label{upperbound}
\frac{s}{\rho + p} = 
\pi l^* \leq
\pi \lambda
\end{eqnarray}
has to be introduced
if the GCEB is to be saved,
apparently being however this exactly provided 
simply by quantum mechanics.
This inequality expresses the fact that 
the requirement of
a lower bound $l^*$ to $\lambda$
translates into an upper bound $\pi \lambda$ to $\frac{s}{\rho + p}$.
Whenever an hypothetical configuration
with assigned $s$, $p$, $\rho$ is considered,
if it presents in some region 
$\frac{s}{\rho + p}$ values exceeding $\pi \lambda$,
it would challenge the GCEB there,
but also, according to the conjecture above,
it would be there simply unphysical
in the sense of quantum-mechanically untenable.
Put another way,
any statistical-mechanical description,
consistent as such with quantum mechanics,
appears to intrinsically require locally
an upper limit $\pi \lambda$ to $\frac{s}{\rho + p}$.

In \cite{Banks},
a connection between entropy and energy densities
(like the ones proven to be sufficient to derive the GCEB)
has been {\it assumed}, 
and has been proposed to be chosen as a constitutive principle
for the construction of a theory of quantum gravity.
The perspective emerging from this section is that,
at our level of description 
(i.e. a statistical-mechanical description for matter
(with entropy, energy and other quantities defined
in the volumes under consideration) and a classical description
for gravity),
the relation between entropy and energy density
to which the GCEB is crucially tied
(relation (\ref{condition})),
is seemingly a straight consequence of quantum mechanics
(through (\ref{condition2}) or (\ref{upperbound})).
It can anyway be considered as a new assumption 
as far as the starting point is chosen to be
a theory in which quantum mechanics is not already in.
In this case, relations (\ref{condition2}) or (\ref{upperbound})
would predict or bring in, it.

\section{The Bekenstein bound derived}

Having $l^*$,
we will prove here the UEB
in a full-gravity setting. 
From the above
this can also be interpreted as implying in particular that the UEB
can actually be derived from the GCEB, i.e. from holography,
without any restriction on the strength of gravity.

The UEB says that
for a physical system
which fits into a sphere of radius $R$
in asymptotically flat 4d spacetime

\begin{eqnarray}\label{UEB}
\frac{S}{E} \leq 2 \pi R,
\end{eqnarray}
where $S$ and $E$ are its entropy and total mass-energy \cite{Bekenstein1}.
Many discussions arose in the past, 
regarding the validity of the bound or its precise meaning 
even for the case of weakly gravitating systems
(see \cite{Bek_Found, Wald_Living} and references therein). 
No controversy seems to survive today 
for the validity of ({\ref{UEB}) if applied
to complete, weakly self-gravitating, isolated systems \cite{Bek_Found}.
However, a completely satisfactory proof 
of the bound even at these conditions,
within the realm of full quantum field theory,
is still lacking, 
not least due to the need to address
the entropy of vacuum fluctuations
(see \cite{Casini} and references therein).
If, 
to obtain a first description of the bound,
this is intended 
as applied first 
only to the fields,
superimposed to vacuum,
which make up the system under consideration
(so that for example $E$ is the energy above the vacuum state),
still a general proof is lacking.
To our knowledge,
the most general proof to date
is, in fact, 
given in terms of 
free fields \cite{SchifferBek}.\footnote{Other statistical-mechanical
arguments which give support to (\ref{UEB}) can be found
in \cite{Khan, Ivanov},
in addition to the original argument \cite{Bekenstein1}.}
If already in Minkowski spacetime some ambiguities
in the definition of $R$ and $E$ have been pointed out \cite{Page},
for curved spacetime it is far from clear
even the meaning they can have in the general case.
At least for spherically-symmetric asymptotically-flat spacetimes however,
most of the inadequacies disappear
and these are the conditions assumed here.

Let us proceed to derive what the notion of $l^*$
implies for the UEB under the stated circumstances,
first with gravity off.
Even if in these circumstances
the GCEB, and thus holography, lose any meaning,
we know that conditions (\ref{condition}) and (\ref{condition2})
have to hold true, and
$l^*$ definition through (\ref{lstar}) remains meaningful.

For any given macroscopic system, 
which we assume at local thermodynamic equilibrium,
we first show that the bound

\begin{eqnarray}\label{weak_gravity}
\frac{S}{E + \int p dV} \leq
\pi R
\end{eqnarray}
holds true,
being $dV$ system's volume element and $R$ the circumscribing radius.
The system, of course, in addition of being
radially inhomogeneus,
can be also in general non-spherically-symmetric, 
as it is clear that for gravity off
the metric, being constant,  is anyhow `spherically symmetric'
around every point.
It can be easily shown, however, that every non-spherically-symmetric
system can always be reduced to a spherically-symmetric one
with a same `circumscribing' radius
and a larger ratio in equation (\ref{weak_gravity}).

Let us consider thus an isolated spherically-symmetric system with radius $R$
and imagine
to ideally perform a partition of it
through a number of concentric spherical shells,
each with thickness $l_i$ 
(with thickness $l_0$ of the central sphere
we intend its diameter $2 r_0$)
small enough that
the shell can be considered at thermal equilibrium with
statistical-mechanical local parameters
constant on it\footnote{A partition like this
is supposed to be always
feasible, on the basis of the notion itself of local equilibrium.}
(actually $l_0$ can be also as large as $2 R$
in the homogeneous case). 
We have 

\begin{eqnarray}\label{weak_gravity2}
\frac{S}{E + \int p dV} =
\frac{\sum S_i}{ \sum \left( E_i + p_i V_i \right) } \leq
\sum \frac{S_i}{E_i + p_i V_i} =
\sum \pi l_i^* \leq
\sum \pi l_i =
\pi R + \pi r_0.
\end{eqnarray}
In the expression above, $l_i^*$ is $l^*$ at the local conditions
of layer $i$.
In (\ref{weak_gravity2})
the first inequality comes from 
$\frac{\sum a_i}{\sum b_i} \leq \sum \frac{a_i}{b_{i}}$ 
if $a_i, b_i \geq 0$;
the second from (\ref{condition}). 

If we allow the spherical shells to be as thin as possible
($l_i$ approaching $\lambda_i$, where $\lambda_i$ is the
typical wavelength (spelled above) on layer $i$
of point-like, or point-like reduced, constituents),
the quantities $\sum S_i$, $\sum E_i$ and $\sum p_i V_i$
in (\ref{weak_gravity2}) still represent, evidently,
total entropy $S$, total energy $E$ and $\int p dV$ respectively.
As long as $R \gg \max \{\lambda_i \}$
as must be always true for macroscopic systems,
we have in (\ref{weak_gravity2}) $r_0 \ll R$
so that (\ref{weak_gravity2}) reduces to (\ref{weak_gravity}). 

When the assumption is made
$p \leq \rho$ 
($p = \rho$ (stiff matter) is the stiffest equation of state compatible
with causality \cite{Zel}),
we have
$S/(2E) = \sum S_i/(2\sum E_i) \leq \sum S_i/\sum (E_i + p_i V_i)$,
so that from (\ref{weak_gravity}),
$S/E \leq 2\pi R$, that is the UEB (\ref{UEB}) is obtained.

From the first inequality in (\ref{weak_gravity2})
we see thus that when $R \gg \max \{ \lambda_i \}$,
the UEB (\ref{UEB}) is satisfied with orders of magnitude
to spare (in the homogeneous case, for example, the first
inequality in (\ref{weak_gravity2}) introduces 
a factor $\sim \frac{R}{\lambda}$).
In general moreover $\lambda$ (or $l_{alw}$) $\gg l^*$
and this contributes further in obtaining
$\frac{S}{E} \ll 2 \pi R$.

These results show 
that to challenge the UEB (\ref{UEB})
systems with size $R \approx \lambda$ must be considered.
The importance of being 

\begin{eqnarray}\label{largeR}
R \gg \lambda
\end{eqnarray} 
for a meaningful analysis of UEB,
has been pointed out in early papers \cite{Bekenstein1,Khan}.
Moreover already \cite{Bekenstein1} argued that a system consisting 
of a photon gas of limiting size  $R \approx \lambda$
comes close to challenge the UEB (\ref{UEB}), in agreement with
what we find.
What is stressed here however is that
condition (\ref{largeR})
on system's size is actually all what is needed
to imply the bound: once this condition is fullfilled,
automatically the UEB is satisfied and nothing is left to be proven.
This relies on inequality (\ref{condition2}), 
which is required by the GCEB,
but, if we trust the conjecture introduced above,
is also a direct expression of quantum mechanics alone.
The linking between the UEB and the uncertainty principle,
contained all the 
arguments above for the bound,
seems to be expressed as straight as possible
by inequality (\ref{condition2}), by which the entire UEB
is summarized: nothing more appears to be needed.

We proceed now to inspect what happens with gravity.
As stated above, 
in search for conditions at which $E$ and $R$ can meaningfully
be defined,
we limit to spherically-symmetric 
asymptotically-flat spacetimes.
We proceed thus to consider a
spherically-symmetric asymptotically-flat spacetime 
from a spherically-symmetric 
system with radius $R$,
assuming to subdivide the system in $N$ concentric spherical shells
each thin enough to be considered at thermal equilibrium,
and we do not pose any constraint on the strength
of gravity.

Following \cite{SWZ}, and in the sense spelled out there,
we restrict our consideration to spacetimes
which admit configurations
with a moment of time symmetry
(in short, these spacetimes do admit
a certain given ``instant of time'' configuration
in  which the extrinsic curvature of 
the corresponding spacelike hypersurface vanishes).
The arguments given 
for this in \cite{SWZ} (strictly, for radiation)
can be summarized in the quest for local extrema of entropy
for given energy. 
Here we argue also that apparently only configurations
belonging to these spacetimes
allow to be completely probed through not expanding lightsheets.

On one hand, in fact,
time-symmetric configurations with $R < 2 M$ 
($M$ is the mass of the system) 
when evolved back into the past inevitably lead
to a white hole \cite{SWZ}.
Assuming white holes cannot exist
(as it is likely to be \cite{Penrose}), 
the instant-of-time-symmetry constraint above
implies the piece of spacetime with $R < 2 M$
should be excluded.
On the other hand,
spherically-symmetric systems inside their own
Schwarzschild radius are not completely covered
by the future-directed ingoing lightsheets
(the only left with non-positive expansion),
the part covered always satisfying the Bousso bound \cite{Bousso}.
Now, the recipe of relating, as in the GCEB, the area of spacelike
2-surfaces with the entropy content of adiacent null hypersurfaces  
appears to be the most convenient approach in view of covariance, 
and of most general applicability, to the entropy bounds.
If, according to this, we assume that the UEB 
(in which $S$ is {\it the} entropy of a given system)
refers to systems completely coverable 
by not expanding lightsheets,\footnote{An extension of the UEB
could however be envisaged, in which $S$ and $E$ are the
entropy and energy not of the whole system but instead of the part
on the future-directed ingoing lightsheet
from the boundary. In these circumstances the UEB would appear to coincide
with the covariant entropy bound.}
we are led, too, to exclude configurations with $R < 2 M$.    

Let us assume $R > 2M$
(with also $r > 2 m(r)$
for any $r < R$ as well,
being $m(r)$ the mass of the part 
of the system enclosed inside the area radius $r$
at a given coordinate time).
Using comoving coordinates
the metric can be written

\begin{eqnarray}\label{metric}
ds^2 =
-e^{2\Phi} dt^2  + e^{2\Lambda} da^2 + 
r^2(d\theta^2 + sin^2\theta d\phi^2)
\end{eqnarray}
with $\Phi$ and $\Lambda$
depending each on $a$ and $t$.
In this expression,
the function $\Lambda$ is related to $m$ \cite{MisnerSharp}
by

\begin{eqnarray}\label{Lambda}
e^\Lambda =
\left( 1 + U^2 - \frac{2m}{r} \right)^{-\frac{1}{2}} 
\frac{\partial r}{\partial a},  
\end{eqnarray}
with $U = e^{-\Phi} \frac{\partial r}{\partial t}$
the comoving proper-time derivative of $r$
and

\begin{eqnarray}\label{m}
m(r) = 
\int_{0}^{r} 4\pi {r^\prime}^2 \rho d{r^\prime} =
\int_0^a \rho \left( 1 + U^2 - \frac{2m(r^\prime)}{r^\prime} \right)^{1/2} dV,
\end{eqnarray}
where $dV = 4\pi {r^\prime}^2 e^\Lambda da^\prime$ is proper volume.
For any given system (thus for assigned local thermodynamic quantities),
and for each $a$, entropy is assigned irrespective of $U$.
From first equality in (\ref{m}) with $r = R$, 
we see $M$ grows with $R$ so that
the configurations which maximize $\frac{S}{M R}$ 
are those with minimum $M$.
Since any $U \neq 0$ implies an increase of $M$,
these configurations correspond to $U = 0$ $\forall a$,
namely they are time-symmetric,
as it could have been expected from what mentioned above.
This means that in our search for a proof of the UEB
we can restrict the consideration to the $U = 0$ case.

We have

\begin{eqnarray}\label{shells}
\frac{S_i}{M_i} =
\frac{s_i}{\rho_i} \ \frac{l_i}{r_i-r_{i-1}} =
\frac{s_i}{\rho_i} \frac{1}{\alpha_i} \leq
\frac{s_i}{\rho_i+p_i} \frac{2}{\alpha_i} =
\pi l_i^* \frac{2}{\alpha_i} \leq
\pi l_i^* \frac{8 m_i}{l_i} \leq
8 \pi m_i,
\end{eqnarray} 
where the index $i$ labels the quantites for the shell $i$ as before
and
$\alpha_i \equiv \left( 1 - \frac{2 m_i}{r_i} \right)^{1/2}$.
Here very small $l_i$ are assumed.
The first inequality comes from causality (according to \cite{Zel});
the second from $\frac{1}{\alpha_i} \leq \frac{1}{\alpha_i^*}$,
where
$\alpha_i^*$ denotes the $\alpha$ at a proper distance $l_i$
from the horizon of a Schwarzschild black hole with mass $m_i$
(and $\frac{1}{\alpha_i^*} = \frac{4 m_i}{l_i}$, see e.g. \cite{MTW});
the third corresponds to (\ref{condition}). 

Using (\ref{shells}) we get
\begin{eqnarray}\label{curved}
\frac{S}{M} =
\frac{\sum S_i}{M} =
\frac{1}{M} \sum \frac{S_i}{M_i} \ M_i \leq
\frac{1}{M} \sum 8 \pi m_i M_i =
\frac{1}{M} \int_0^M 8 \pi m dm = 
4 \pi M \leq 
2 \pi R,
\end{eqnarray}
that is the UEB (\ref{UEB}) with $E = M$.

In (\ref{shells}) the second inequality turns into an equality
when $2 m_i$ and $r_i$ are just close enough that,
if $m_i$ is put into the sphere ``$i-1$'',
a black hole is formed.
This happens when  $l_i \alpha_i = r_i - 2 m_i$,
or $l_i = r_i \alpha_i$.
If $r_i$ becomes closer to $2 m_i$ the inequality 
would be violated for the assigned $l_i$,
but we can choose a smaller $l_i$, such that
the equality holds true.
Continuing this process
we reach a situation which demands for $l_i = \lambda_i$ or smaller.
We meet here a crucial point.
We have to face how the notion of black hole should be modified
in a context in which
the quantum nature of matter is taken into account.
A macroscopic state with $r_i \alpha_i < \lambda_i$
seems impossible to be conceived
due to the quantum uncertainty in the position of the constituent particles.
In such a state in fact,
as a fraction of the wavelength of the most external
particles extends to $r < 2 m_i$, 
there would be a non-vanishing probability
to have an horizon.
As a consequence, other particles,
also due to their own random motion, 
could find themselves
inside the just considered horizon, which would have to be larger,
and so on, so that the (logical) process would end only with 
all particles inside.
This is a quantum-mechanical argument regarding the macroscopic state, 
irrespective of the motion of the particles 
in the gravitational field. 
This suggests we have to jump directly from the macroscopic state with
$r_i \alpha_i = \lambda_i$
to the black hole. 
According to this, the second inequality in (\ref{shells}) is always
satisfied and is just attained when a black hole is formed.

Relation $\lambda_i \leq r_i \alpha_i$, indeed,
turns out to be the kind of inequality found in \cite{SWZ}
(cf. eqs. (2) and (52) there) considering
the physical feasibility of configurations
arbitrarily close to the Schwarzschild radius,
given the intrinsic quantum-mechanical size of
constituent particles.\footnote{We find 
that the violation (at $\sim$ Planck scale)
of the Bousso bound, and of the UEB, 
found in \cite{Gersl},
appears to be accompanied by a violation 
of the condition $\lambda_i \leq r_i \alpha_i$,
in that, 
at the circumstances considered in \cite{Gersl}
as challenging the bound,
the radius of the system turns out to be smaller than 
$\lambda/{\sqrt{1-\frac{2M}{R}}}$.}
Looking at (\ref{shells})
the attaining of the UEB limit by not collapsed bodies
is possible only with
stiff matter.\footnote{For gravitating systems,
the peculiarity of stiff matter
can be seen also in the
approximate area scaling of entropy
of stiff-matter balls
(beyond a certain radius depending on central pressure)
\cite{BanksFischlerAl} (see also \cite{NoExt}).}
We notice a benefit from gravity.
In flat spacetime, we have seen,
the configurations able in principle to attain
the UEB (\ref{UEB}) are from a statistical-mechanical
point of view rather peculiar,
as they correspond to systems with $R \approx \lambda$.
Thanks to gravity,
we can consider equivalent systems
(that is with a same $\frac{S/E}{R}$ just attaining the UEB)
at any radius,
with sound statistical-mechanical meaning.
%

If we consider a plane layer thin enough
to allow for statistical-mechanical quantities
to be considered constant in it,
from (\ref{condition})-(\ref{lstar}) we have
$S \leq \pi (\rho+p) A l \cdot l$, with $A$ the surface area
of the layer and $l$ the proper thickness.
For weak-field conditions, 
this expression becomes

\begin{eqnarray}\label{genBek}
S \leq \pi \left( E + p A l \right) \cdot l,
\end{eqnarray}
with $E$ the total mass-energy of the layer of matter.
When $p \ll \rho$ inequality (\ref{genBek})
can be written as

\begin{eqnarray}\label{BoussogenBek}
S \leq \pi E l,
\end{eqnarray} 
which is the expression of the ``generalized'' Bekenstein bound 
given
in \cite{Bousso2003} (eq. (15) there;
with $l$ intended as the smallest size
of the body)
when applied to this layer,
and which is tighter, 
for non-spherically-symmetric bodies as in the present case,
than the original one of Bekenstein.
For non-negligible $p$, the general bound to the entropy on the layer
can be expressed as

\begin{eqnarray}\label{gengenBek}
S \leq 2\pi E l,
\end{eqnarray}
since $p\leq \rho$ if we require causality \cite{Zel}.
For spherical shells, 
inequalities (\ref{genBek}), (\ref{BoussogenBek}) and (\ref{gengenBek})
remain unaffected, with $l$ still denoting layer's thickness 
(and thus inequality (\ref{BoussogenBek}) becomes here
much stronger than what apparently 
implied by \cite{Bousso2003}, 
which refers to the smallest size of the body,
that is here the diameter $2 R$ of the shell). 
In the limit of $l$ approaching the radius $R$,
inequality (\ref{gengenBek}) becomes the UEB (\ref{UEB}).

Let us consider a plane layer of matter
just attaining the bound in inequality (\ref{gengenBek})
(thus with $l = \lambda = l^*$ and $\rho = p$)
with assigned area.
Imagine to increase the energy per particle,
at the same time reducing the thickness of the layer
to coincide, still, with particle's new $\lambda$,
so that the entropy in the layer is always at the limit.
Let us imagine also, we do this while lowering energy density
of the amount needed to leave total energy constant.
No end can be, seemingly, envisaged for such a procedure
from a quantum-mechanical standpoint,
so that inequality (\ref{gengenBek}) can hold true 
without any limit to the smallness of $l$. 
One additional information, 
not coming from quantum mechanics, 
must be introduced, 
namely the concept of gravitational collapse,
to understand that
beyond a certain limiting energy per particle (the Planck scale)
particles behave as, let say, black holes, 
and any further increase
of particle's energy necessarily brings to an increase in the
thickness of the layer instead of a decrease
(reflecting this, a modification of Heisenberg's relations
at Planck scale) if the layer has not to become thinner
than particle's size. And in this new scenario inequality (\ref{gengenBek})
will hold true anyway.

We notice that
a decreasing $\lambda$
implies
total entropy $S \propto l = \lambda$
decreasing as $\lambda$.
Entropy per unit area $\sigma$
in the region where the particle is localized, 
goes, however, as $\frac{S}{\lambda^2}$
and thus as $\frac{1}{\lambda}$,
that is it increases when $\lambda$ decreases.
When $\lambda$ approaches the Planck scale,
$\sigma$ reaches its maximum value which,
as the 2-surface
near the particle can no longer be considered
plane, is $\sigma_{max} = \left( \frac{S}{A} \right)_{bh} = \frac{1}{4}$;
any further increase in particle's energy cannot give
any variation to this value of $\sigma$.
The maximum entropy per unit area of a layer with thickness
equal to constituent particle's size is thus given by
$\sigma_{max} = \frac{1}{4} \frac{1}{l_p^2}$,
being $l_p$ the Planck length.
From inequality (\ref{gengenBek}) 
taken alone, $\sigma$ could have grown instead without
any limit ($\sigma \sim \frac{1}{\lambda}$).
We see that if quantum mechanics means setting a limit on {\it specific}
entropy (entropy per unit energy),
gravity does not produce any change to this limit 
but amounts to set, from it,
a limit on {\it absolute} entropy per unit area.

In conclusion,
in this section we have given a proof of the UEB 
for any strength
of the gravitational interaction using holography in its most general
semiclassical formulation as given by the GCEB.
This has been done using the $l^*$ concept.
In the following section the connection of the UEB
with the $\eta/s$ bound and with the proliferation of species
problem is considered,
still from $l^*$ standpoint.

\section{Some related issues}

In \cite{FBB},
the UEB (\ref{UEB}) 
is used to put a local constraint on
$s/\rho$ values for a fluid,

\begin{eqnarray}\label{FBB}
\frac{s}{\rho} \leq 2\pi L,
\end{eqnarray}
in terms of the correlation length $L$ of the material medium,
arguing that
this is the minimum size for a parcel of fluid 
to have a consistent fluid description
(i.e. with the usual equations of dissipative hydrodynamics).
In the perspective we have presented here,
we obtain (see (\ref{gengenBek}))

\begin{eqnarray}\label{local}
\frac{s}{\rho} \leq 2\pi l^* \leq 2\pi \lambda,
\end{eqnarray}
with $l^*$ given by (\ref{lstar}).
Here the first inequality comes simply
from $p \leq \rho$.
Bound (\ref{local}) is, in general, much stronger
than bound (\ref{FBB}). If we think for example
to an ordinary gas, $L$ is given by the mean free path,
and is much larger
than the quantum uncertainty $\lambda$
on the positions of the molecules.  
The fact that (\ref{condition2}) is in general a strong inequality
(whichever is the state of the material medium,
i.e. if gaseous, liquid or solid) gives anyway
that bound (\ref{local}) is in general satisfied to spare.  

For an assigned material medium
a bound of the kind (\ref{FBB}) cannot be the final word on $s/\rho$.
$L$, in fact, changes in general
with $s$ and $\rho$, even when intensive thermodynamic parameters
and $s/\rho$ ratio are held fixed. 
Assuming thus that, 
while holding the ratio $s/\rho$ fixed,
conditions can always be envisaged
for which the correlation length is reduced to the particle's 
$\lambda$ corresponding to the assigned intensive parameters
(below $\lambda$ we cannot go, due to quantum correlation),
starting from (\ref{FBB})
we are brought eventually to (\ref{local}).

Some consequences can be drawn from this, concerning
the viscosity to entropy ratio $\eta/s$.
For nonrelativistic systems,
assuming $\eta = \frac{1}{3} L \rho a$ 
($a$ is thermal velocity) \cite{Huang},    
we get

\begin{eqnarray}\label{visco_nrel}
\frac{\eta}{s} =
\frac{1}{3} \ \frac{L \rho a}{s} \simeq
\frac{1}{3} \ L \ \frac{\rho + p}{s} a =
\frac{1}{3\pi} \ \frac{L}{l^*} a,
\end{eqnarray}
where last equality comes from (\ref{lstar}).
We can consider now another configuration of the same system
which has minimum $\eta/s$ for the assigned intensive parameters. 
To this aim we act as described above on $\rho$, $s$ and $p$,
while keeping the intensive variables unchanged 
(and $\frac{\rho+p}{s}$ and $\lambda$),
till obtaining $L$ approaches its quantum-mechanically 
limiting value $\lambda$, if feasible.
We have

\begin{eqnarray}\label{visco_min}
\left( \frac{\eta}{s} \right)_{min} =
\frac{1}{3\pi} \ \frac{\lambda}{l^*} a.
\end{eqnarray}
Now if, 
to try to understand what is going on,
we consider a Boltzmann gas,
we have (from \cite{Pesci3})
$\frac{\lambda}{l^*} \propto 
\sqrt{\frac{m}{T}} \propto
1/a$ (being $\frac{1}{3} m a^2 = T$ \cite{Huang}),
with $m$ the mass of constituent particles,
and this means that in the expression (\ref{visco_min}) no dependence
on $a$ is left.
Using the expression for $\frac{\lambda}{l^*}$ given in \cite{Pesci3}
(with the parameter $\chi \ll 1$ there, since we are assuming
$L$, the mean free path, is approaching $\lambda$)
we obtain
$\left( \frac{\eta}{s} \right)_{min} \approx 0.7$
(for $g = 2$, being $g$ the number of degrees of freedom per particle).
Thus we notice that:\\
i) the KSS bound \cite{KSS}
$\eta/s \geq 1/4\pi$
seems satisfied even if $a \ll 1$ (see also \cite{FBB});\\
ii) considering (along the lines of \cite{FBB})
the KSS bound also as an entropy bound ($s \leq 4\pi \eta$),
(nonrelativistic) systems 
which are far away, by many orders of magnitude,
from the attaining of (\ref{local}) 
($s \ll 2\pi \rho \lambda$),
can anyway be near to saturate the KSS bound,
so that the latter appears for such systems much tighter
than (\ref{local}) intended as an entropy bound.

For ultrarelativistic systems,
if we refer to a fluid consisting of
particles whose statistical equilibrium is determined completely
by collisions through radiation quanta (being, these, photons
or neutrinos, or gluons), 
assuming that 
$\eta \approx \frac{1}{3} \tau \rho_{\gamma}$ \cite{Misner},
where $\tau$ is the average time for a quantum to collide
and $\rho_{\gamma}$ is the energy density of radiation,
and that the contribution to total entropy from radiation is dominant,
we get

\begin{eqnarray}\label{visco_rel}
\frac{\eta}{s} \approx
\frac{1}{3} \tau \frac{\rho_{\gamma}}{s_{\gamma}} =
\frac{1}{4} \tau \frac{\rho_{\gamma} + p_{\gamma}}{s_{\gamma}} =
\frac{1}{4\pi} \frac{\tau}{l_\gamma^*} =
\frac{1}{4\pi} \frac{L}{l_\gamma^*}
\end{eqnarray}
where the quantities with subscript $\gamma$ refer
to radiation and as correlation length $L$
we intend the average path for a quantum to collide.
At conditions such that $L$ 
has its limiting value (when this is allowed),
namely the wavelength itself $\lambda_\gamma$ of the quantum
of radiation, the ratio $\frac{\eta}{s}$ has its minimum
given by

\begin{eqnarray}\label{visco_rel_min}
\left( \frac{\eta}{s} \right)_{min} \approx
\frac{1}{4\pi} \frac{\lambda_\gamma}{l_\gamma^*} =
\frac{1}{4\pi},
\end{eqnarray}
which just corresponds to attain,
through conventional quantum mechanical arguments, 
the string theoretical KSS bound.
Here, last equality follows from the saturation of 
the bound (\ref{condition2}) for ultrarelativistic fluids
(with $\mu = 0$) \cite{Pesci3}.
As far as the mentioned conditions 
can be considered conceivable
in the quark-gluon plasma,
what we have just obtained could be considered also
as a back-of-the-envelope description
of the very low values of $\frac{\eta}{s}$ \cite{QGP,QGP_theory}
experimentally found for it (see also \cite{FBB}).\footnote{Regarding
this point of the role that ordinary quantum mechanics can play
in providing a very low $\eta/s$, see also \cite{Gyulassy}.}
We see that the viscosity bound,
as the UEB, can be thought as coming 
from $l^*$ concept (inequality (\ref{condition2})). 

Some consequences of the results we have presented,
can also be drawn 
concerning
the so-called proliferation of species problem.
The problem \cite{SWZ,species,FMW,Wald_Living}
(see also \cite{Bousso_review} and references therein)
is in that if one allows
for an unlimited proliferation of particle species,
one should expect the entropy in a box at fixed energy to challenge
the entropy limit envisaged by the UEB.
Indeed, also the statistical-mechanical main proof of the UEB
as given in \cite{SchifferBek}, refers to a limited number 
(a very liberal limit, actually)  
of non-interacting fields. 

Many ways to face 
this problem have been proposed in the years
(\cite{Bek_Found} and \cite{Bousso_review}, and references therein),
and we refer the reader to \cite{Bek_Found} 
for a quite recent review on this.
$N$ could be intrinsically required to be limited
for physics to be consistent
(see \cite{BrusteinAl}, concerning the stability
of the quantum vacuum itself).
However, even if $N$ could be allowed to be arbitrary,
something which is assumed neglibible in present arguments 
in favour of the UEB, and in its counter-arguments,
could turn out however to be essential when $N$ is large,
so that the UEB could still
be satisfied \cite{Bek_Found}.
As a general point we could say that
the UEB could be true
whatever the consistency of physics demands to $N$, 
i.e. whether it must be limited or can be arbitrary
(see also \cite{Casini}).

This is also what is obtained in our approach.
Condition (\ref{condition2})
translates in this case
to the request that some representative value of the collection of 
$\lambda$'s (corresponding to the different species) 
be not smaller than $l^*$, as defined by (\ref{lstar}).
This looks quite
insensitive to the number of species.
If for example we consider
$N$ copies of the electromagnetic field at a given $T$,
with $N$ arbitrary,
denoting with $\rho_N$, $s_N$, ... the quantities referring
to the system with $N$ copies and with
$\rho_1$, $s_1$, ... the quantities for the case of one copy, 
we get
$\rho_N = N b T^4$ and
$s_N = \frac{4}{3} N b T^3$ with $b$ a constant,
so that $l^*_N  = l^*_1$,
and condition (\ref{condition2}), 
which is in this case $\lambda \geq l^*_N$, 
is satisfied,
attained to be more precise, as in the case with only one copy.
For the chosen circumstances however also the ratio $S/E$ is insensitive
to $N$.

If the $N$ copies have to be taken with total energy fixed,
in the $N$-copy case 
the temperature must be different,
$T \rightarrow T^\prime$,
with $T^\prime$ satisfying $N {T^\prime}^4 = T^4$.
In this case
$\rho_N = \rho_1$
and
$s_N = \frac{4}{3} N b {T^\prime}^3 =
N^{1/4} s_1$,
so that  
$l^*_N  = N^{1/4} l^*_1$.
Also photon wavelengths however must be different,
$\lambda = \frac{1}{\pi T} \rightarrow 
\lambda^\prime = \frac{1}{\pi T^\prime}$
and thus, still,
condition (\ref{condition2}),
namely $\lambda^\prime \geq l^*_N$ if we have $N$ copies,
is satisfied (attained) as for the one-copy case.
The proliferation of species, thus, could be not a problem
for condition (\ref{condition2}), too.
It would hold true regardless of the properties of $N$,
being it bounded or arbitrary.

In conclusion,
in this paper a proof of the UEB, the universal entropy bound, 
for any strength
of gravity has been given, using holography in its most general
semiclassical formulation as expressed by the GCEB,
the generalized covariant entropy bound.
A modified formulation and a proof of the UEB
for plane layers and for spherical shells has been also proposed. 
What has been used in the proofs is only
the $l^*$ concept 
(equation (\ref{lstar}) and inequality 
(\ref{condition}) or (\ref{condition2})), 
a concept essential for the GCEB
but fully meaningful also in a world without gravity.
In such a world the GCEB would lose any meaning,
contrary to the UEB.
The emerging perspective is thus
that both the GCEB and the UEB arise from the $l^*$
concept, being the former obtained when gravity is turned on
and the latter straight from $l^*$ without any reference
to gravity.
Some consequences from $l^*$ standpoint 
as for the $\eta/s$ bound and the proliferation of species problem
have also been drawn.

\vspace {0.5cm}
{\bf NOTE ADDED IN PROOF}

When the present manuscript was (almost) completed
new results have been published dealing
with the relation between the viscosity bound and the
generalized second law of thermodynamics \cite{Hod2}.
The KSS bound is derived in them 
using an approach completely different from that presented here,
but still relying on the existence and value of
the minimum scale $l^*$.

\end{document}